# Engineering shallow spins in diamond with nitrogen delta-doping

Kenichi Ohno, F. Joseph Heremans, Lee C. Bassett, Bryan A. Myers, David M. Toyli, Ania C. Bleszynski Jayich, Christopher J. Palmstrøm, and David D. Awschalom

*Center for Spintronics and Quantum Computation, University of California, Santa Barbara, California 93106, USA*

**Abstract**

We demonstrate nanometer-precision depth control of nitrogen-vacancy (NV) center creation near the surface of synthetic diamond using an *in situ* nitrogen delta-doping technique during plasma-enhanced chemical vapor deposition. Despite their proximity to the surface, doped NV centers with depths ($d$) ranging from 5 - 100 nm display long spin coherence times, $T_2 > 100$ µs at $d = 5$ nm and $T_2 > 600$ µs at $d \geq 50$ nm. The consistently long spin coherence observed in such shallow NV centers enables applications such as atomic-scale external spin sensing and hybrid quantum architectures.

The nitrogen-vacancy (NV) center is an individually addressable electronic spin in diamond that can be initialized and read out optically[1] and coherently controlled with microwaves[2] at room temperature. Together with the exceptionally long spin coherence times ($T_2$) of NV centers in isotopically engineered materials,[3] these features enable the detection of weak coupling between single NV centers and other quantum degrees of freedom, motivating a wide variety of applications. Demonstrations of NV center spins as nanoscale probes include magnetometry,[4-6] magnetic imaging,[7,8] electric-field sensing,[9] and thermometry.[10] For quantum information schemes they have been employed in hybrid quantum systems,[11-13] long-lived quantum memories[14,15] and the controlled manipulation of quantum registers.[16-19] Still, certain applications, such as proposals for single electron and nuclear spin imaging and external spin entanglement,[20,21] are limited by the weak magnetic dipole interaction, with a coupling strength that decreases with distance ($r$) as $1/r^3$. This requires that NV centers be located within a few nanometers of the surface. The very low concentrations of NV centers in bulk ultrapure diamond make near-surface spins extremely rare, necessitating an artificial means for their creation.

Engineering surface-proximate NV centers with long spin coherence times remains a challenge. Most efforts have used nitrogen implantation, which suffers from a large depth dispersion even at relatively low acceleration energies due to ion straggling and channeling effects[22] (e. g., nominally 10-nm-deep NV centers implanted at 6 keV have dispersion >10 nm[8]). While channeling effects diminish for very low energy implantations,[23] the resulting implanted spin coherence times are generally short and widely dispersed[24] even in isotopically purified substrates,[16] presumably due to crystal damage intrinsic to the impact of accelerated ions. As an alternate approach, *in situ* nitrogen doping techniques show promise in this regard. Encouragingly, growth-incorporated NV centers with millisecond-long $T_2$ times have recently been located within the top 100 nm of isotopically purified $^{12}$C diamond,[25] although the distribution of depths is still too large for many sensing applications.

Here, we report a growth technique using delta-doping of nitrogen for nanometer-scale depth control of NV center creation near the surface. By reducing the growth rate to

~0.1 nm/min, a thin nitrogen-doped layer (1-2 nm) is created by the controlled introduction of $N_2$ gas during plasma-enhanced chemical vapor deposition (PE-CVD) of high-quality, single-crystal diamond. The vacancies needed to form NV centers are created separately from the nitrogen incorporation by *ex situ* electron irradiation and subsequent annealing,[26] causing far less damage in the nitrogen-doped layer than ion implantation techniques.[24] Therefore, we are able to maintain the high crystal quality of PE-CVD-grown diamond to provide both depth localization and reproducible, long spin coherence of artificially created NV centers.

We have optimized the PE-CVD diamond growth process for high-precision layer structures, using a temperature of 800 °C, pressure of 25 Torr, $H_2$ flow of 400 sccm, $CH_4$ flow of 0.1 sccm, and microwave power of 750 W[27] to grow on top of commercially available *Element Six* electronic grade (100) substrates. These growth conditions produce a growth rate of 8.2 ± 3.2 nm / hour, calibrated by secondary ion mass spectrometry (SIMS) on selected samples. We grow with isotopically purified $^{12}CH_4$ (99.999 %) gas in order to mitigate decoherence caused by $^{13}C$ nuclear spin fluctuations. In addition, our technique allows us to engineer specific nuclear spin environments in isotopically layered diamond structures by switching between purified $^{12}CH_4$ and $^{13}CH_4$ (99.99 %) source gases.

Nitrogen is incorporated during growth by using a mass flow controller to introduce $^{15}N_2$ gas with specified timing and duration. We use isotopically purified $^{15}N_2$ (> 98 %) to distinguish NV centers formed within the doped region from those due to naturally occurring $^{14}N$ impurities (99.6% isotopic abundance, <5 ppb) in the substrate.[28] Varying the $N_2$ flow from 10 to 50 sccm resulted in a monotonic increase of doped nitrogen concentration from (0.8 ± 0.6) x $10^{16}$ cm$^{-3}$ to (3.1 ± 0.3) x $10^{16}$ cm$^{-3}$ as detected by SIMS; for a 5 sccm $N_2$ flow the concentration was below the SIMS noise level of 0.6 x $10^{16}$ cm$^{-3}$.[29] Given the $N_2/CH_4$ flow ratio, the unexpectedly low nitrogen incorporation could be attributed to inefficient activation of $N_2$ due to the low microwave power or to a dominant hydrogen etching effect[27] removing most of the nitrogen adatoms. Contrary to the case of high microwave power density,[30-32] no growth rate enhancement due to the introduction of nitrogen gas was observed, allowing for precise thickness control of the nitrogen doped layer. Vacancies are created via electron irradiation ($10^{14}$ cm$^{-2}$ dose at 2 MeV) and the samples are annealed in forming gas at 850 °C for 2 hours to form NV centers. The surface is oxygen terminated by an acid process (H$_2$SO$_4$: HNO$_3$: HClO$_4$ = 1:1:1 at 200 °C for 30 min.) in order to stabilize the NV centers in the negatively charged state.[33,34]

The spin properties of NV centers in our samples were investigated using a home-built confocal optical microscope[35] to measure the spin-dependent photoluminescence intensity ($I_{PL}$) under optical excitation at 532 nm, while applying microwaves to manipulate the ground-state spin. We distinguish doped NV centers from those in the substrate by the hyperfine coupling between their electronic spin and intrinsic nitrogen nuclear spin, measured using low-power continuous-wave electron spin resonance (CW-ESR) spectroscopy. Figure 1(a) shows the CW-ESR spectrum of an NV center in the substrate, displaying a three-fold hyperfine splitting indicative of the $^{14}$N nuclear spin ($I$ = 1). In contrast, doped $^{15}$NV centers show a two-fold splitting (Fig. 1(b)) induced by the hyperfine coupling to a $^{15}$N nucleus ($I$ = 1/2). Since the natural abundance of $^{15}$N is only 0.4%, isotopic discrimination in this manner provides a reliable signature that a given NV center originated from the $^{15}$N-doped layer.

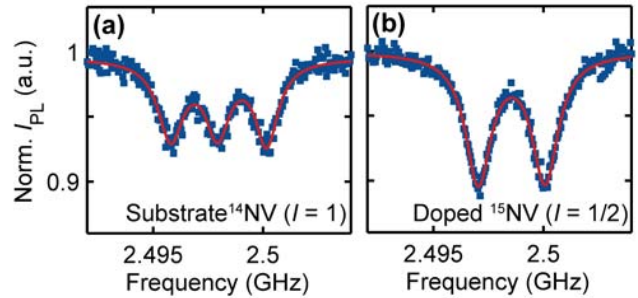

**Figure 1:** (Color) CW-ESR spectra of a substrate $^{14}$NV center (a) and a doped $^{15}$NV center (b). The curves show three-, and two-Lorentizan fits to the respective data.

In order to constrain the depth distribution of doped NV centers, we created the isotopically-layered diamond structure shown schematically in Fig. 2(a) superposed on a SIMS measurement of the $^{13}$C concentration. A $^{15}$N-doped layer ($^{15}$N$_2$ flow: 10 sccm, thickness $t$ = 2 nm) was formed at the center of a 15-nm-thick $^{12}$C-enriched diamond layer sandwiched between two $^{13}$C-enriched diamond layers. We use the nanometer-scale coupling between NV center spins and nearby $^{13}$C nuclei to probe the local environment of doped NV centers, as detected by the electron spin echo envelope modulation (ESEEM)[36] of coherence in a Hahn echo microwave pulse sequence ($\pi/2 - \tau - \pi - \tau - \pi/2$, with two free precession intervals of equal duration $\tau$).[2] Couplings between an NV center and different nuclear spin species can be distinguished by the frequency components of an ESEEM measurement. The $^{15}$N nuclear spin of doped NV centers induces modulations at two frequencies, $f_{m_s}^{15N}$, corresponding to the $^{15}$N Larmor precession frequencies for electron spin sublevels $m_s$ = 0 and -1. For the magnetic fields used in this work, $f_{-1}^{15N} \approx 3$ MHz is the hyperfine coupling frequency, and $f_0^{15N}$ is approximately proportional to the component of the external magnetic field perpendicular to the NV-center symmetry axis.[37,29] In addition, weak coupling to $^{13}$C nuclear spins induces a third frequency, $f^{13C}$, corresponding to the bare Larmor precession of $^{13}$C nuclei in the external field.[38,39] The $^{12}$C atoms do not produce oscillations because they have no nuclear spin, nor does the $^{14}$N nuclear spin in a $^{14}$NV due to its large quadrupolar splitting.[38] Since NV centers couple only to $^{13}$C nuclei within a distance of a few nanometers,[29] the ESEEM measurement reveals whether a doped NV center is confined to the 15-nm-thick $^{12}$C layer or dispersed into the $^{13}$C layers.

Figure 2(b) shows the ESEEM signal of a $^{14}$NV center ~15 μm deep within the substrate, demonstrating collapses and revivals with $f^{13C}$ = 145 ± 1 kHz, corresponding to $^{13}$C precession in an applied magnetic field of 136 ± 1 G. These oscillations are induced by $^{13}$C nuclear spins naturally abundant (1.1 %) in the substrate. In contrast, Fig. 2(c) shows a measurement at the same magnetic field of a doped NV center, clearly displaying two frequency components, $f_{-1}^{15N}$ = 3.15 ± 0.01 MHz and $f_0^{15N}$ = 179 ± 3 kHz (all uncertainties from fits are quoted at 95 % confidence). These are the expected hyperfine coupling frequency and the $^{15}$N Larmor frequency corresponding to a perpendicular field of 30 G, respectively, and are clearly distinct from the $^{13}$C-induced modulation frequency.

Furthermore, the off-axis coupling $f_0^{15N}$ effectively produces periodic *revivals* of coherence by modulating the depth of oscillations at the hyperfine frequency $f_{-1}^{15N}$, while any $^{13}$C coupling would result in *collapses* not observed in Fig. 2(c). Therefore, we conclude that this doped NV center must be in the thin $^{12}$C-enriched layer.

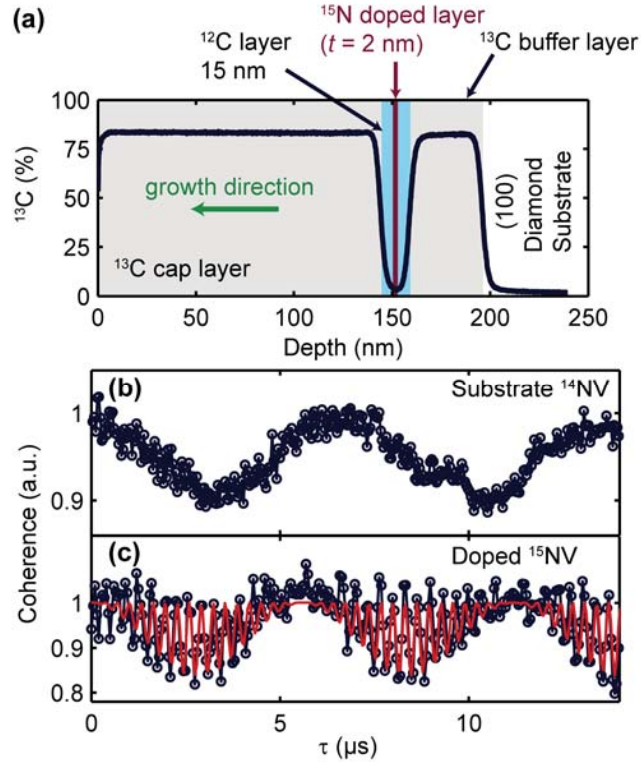

**Figure 2:** (Color) (a) $^{13}$C atomic percentage, inferred from $^{13}$C and $^{12}$C SIMS measurements of the $^{13}$C/$^{12}$C/$^{13}$C isotopically layered structure. A schematic of the sample structure is superposed on the SIMS data, showing the $^{15}$N doped layer (thickness 2 nm) as a red line at the center of the 15nm-thick $^{12}$C layer. (b)(c) ESEEM measurements of a substrate $^{14}$NV center (b) and doped $^{15}$NV center (c). The curve in (c) is a fit according to the ESEEM model[29] with frequency components $f_0^{15N}$ and $f_{-1}^{15N}$.

Of 29 doped NV centers we located in this $^{13}$C/$^{12}$C/$^{13}$C structure, 23 showed no evidence of coupling to $^{13}$C, and so must be fully within the $^{12}$C layer. We calculate[29] that NV centers located more than ~ 3 nm from the $^{13}$C-enriched buffer or cap layers will not display detectable coupling to $^{13}$C spins, implying a maximum depth dispersion for doped NV centers, given that ~ 80 % showed no measurable coupling, of $\sigma$ ~ 4 nm. The remaining 6 NV centers showed modulation at all three frequencies $f^{13C}$, $f_0^{15N}$, and $f_{-1}^{15N}$. Of these, however, the shortest observed $T_2$ was 70 ± 10 μs while the other 5 exhibited $T_2$ > 250 μs. This is much longer than the expected $T_2$ for NV centers in an isotopically enriched environment of $^{13}$C > 80 % (~ 8 μs assuming $T_2$ is inversely proportional to the $^{13}$C concentration),[40,41] suggesting that these 6 NV centers are still located well within the thin $^{12}$C layer. Since the echo collapses observed in ~ 20 % of doped NV centers could result from residual $^{13}$C impurities in the $^{12}$C region from the buffer layer growth rather than coupling to the $^{13}$C layers, the dispersion could still be as small as the 2-nm layer thickness inferred from the growth rate.

To systematically study the effects of fabricating NV centers close to the surface, we grew a series of samples with the structure shown schematically in Fig. 3(a). Nitrogen is delta-doped in a single layer of $^{12}$C diamond film grown on top of (100) diamond. In order to mitigate spin decoherence caused by substrate $^{13}$C nuclear spin fluctuations, we first grew a $^{12}$C buffer layer of 32 ± 12 nm thick, followed by the $^{15}$N-doped layer (thickness: $t$) and finally a $^{12}$C cap layer of thickness $c$ = 4 - 100 nm. The average depths, $d$, of NV centers in our samples are $d$ = 5.1 ± 2.6, 21.0 ± 8.7, 52 ± 21, and 103 ± 41 nm, where $d = c + t/2$, and thicknesses and corresponding 1$\sigma$ uncertainties are computed from the SIMS-calibrated growth rate. For the samples with $d \geq$ 20 nm, the $^{15}$N doping condition ($^{15}$N$_2$ flow: 10 sccm, for 10 minutes) results in a doped layer thickness of $t$ = 1.3 ± 0.5 nm. For the $d$ = 5 nm sample, we used an increased $^{15}$N$_2$ flow (30 sccm for 15 minutes), producing a doped layer with $t$ = 2.0 ± 0.8 nm. In contrast to the consistent doped NV center density we observed in samples with $d \geq$ 50nm (~2-3 x 10$^{13}$cm$^{-3}$), the density in some samples with $d$ < 40 nm was very low (< 10$^{12}$ cm$^{-3}$). For this reason we increased the $^{15}$N concentration in the $d$ = 5 nm sample. The increased N concentration could compensate for decreased NV-center creation efficiency near the surface, possibly due to mobile vacancies annihilating at the surface.[42] Also, the added valence electrons from additional N could stabilize the negatively charged state of the NV center.[33,34]

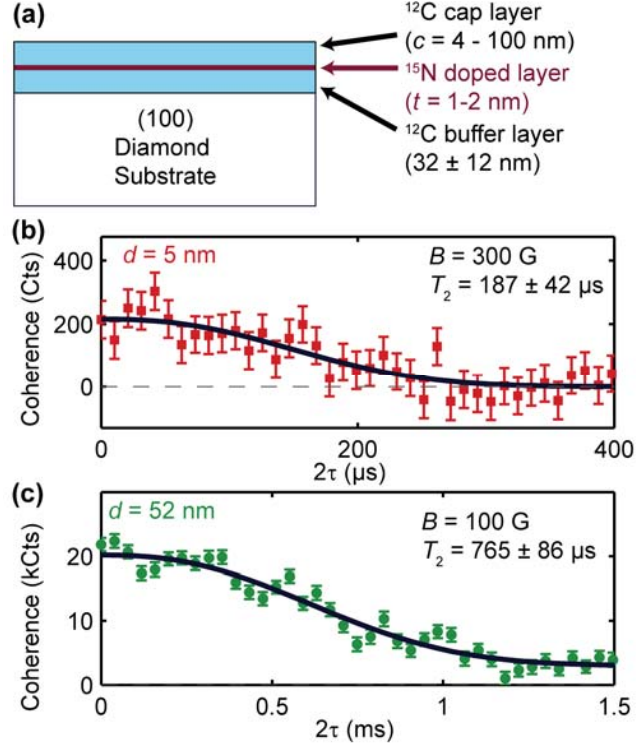

**Figure 3:** (Color) (a) Schematic of $^{15}$N delta-doped $^{12}$C films with varying cap layer thickness, $c$. (b) Hahn echo coherence decay of doped NV centers at $d = 5$ nm (top) and 52 nm (bottom). The curves are fits according to equation (1) in the main text. Uncertainties in $T_2$ are quoted at 95 % confidence.

As before, we identified doped $^{15}$NV centers in each sample using their $^{15}$N hyperfine signature in CW-ESR spectroscopy. We measured $T_2$ by performing a Hahn echo sequence, choosing free precession times $\tau$ equal to multiples of the slow revival period, $1/f_0^{^{15}N}$, to isolate the decay of the coherence envelope from the fast modulations at frequency $f_{-1}^{^{15}N}$.[29] Figure 3(b) shows the Hahn echo coherence decay of doped NV centers at $d = 5$ (top) and 52 nm (bottom), respectively, with fits of the form

$$F(\tau) = A_0 \exp\left[-(2\tau/T_2)^n\right] + y_0, \qquad (1)$$

where the exponent $n$ is allowed to vary in addition to fitting parameters $A_0$, $y_0$, and $T_2$. Despite the relatively high nitrogen doping concentration (0.8-3.1 x $10^{16}$ cm$^{-3}$) and proximity to the surface, we observed several NV centers with $T_2 > 100$ μs at $d = 5$ nm and $T_2 > 600$ μs at $d = 52$ nm. Since some decoherence can be mitigated through dynamical decoupling,[43-45] $T_2$ represents a lower limit on the achievable coherence time. Still, $T_2$ provides a good benchmark for judging the quality of the material.

Figure 4(a) summarizes the results of coherence measurements for all doped NV centers studied in these samples, demonstrating repeatable $T_2$ within each sample. In contrast, ion implantation techniques typically produce inconsistent results, with a non-negligible probability (up to 40 %) of creating NV centers with very short $T_2$ (< 10 μs).[24] Since $^{13}$C is absent in our samples, it is likely that the relatively high nitrogen concentration in the doped layer limits the coherence times in these samples. Doped NV centers with $d \geq 50$ nm showed shorter $T_2$ than for NV centers in bulk isotopically pure diamond[3] and these observed values are consistent for our nitrogen concentration level, assuming $T_2$ is inversely proportional to the total paramagnetic impurity concentration.[46] We therefore hope to further improve the spin coherence time by reducing the nitrogen doping concentration. While the increased nitrogen concentration in the $d = 5$ nm sample probably contributes to the reduction of $T_2$, surface proximity effects might additionally reduce $T_2$ in the $d = 5$ and 21 nm samples; however, further study is necessary to rule out other possibilities such as variations in substrate quality and nitrogen incorporation.

In order to motivate using these shallowly doped NV centers for surface spin sensing applications, we show in Fig. 4(b) the sensitivity to an external magnetic dipole[29] as a function of distance ($r$) from the NV center, for the highest-coherence NV centers in the $d = 5$ and 21 nm samples. We assume the external spin can be initialized, oriented to maximize the projection of its dipole magnetic field along the NV axis, and driven resonantly to enable ac magnetometry techniques.[4,5,11,47] Despite shorter $T_2$, the NV center at $d = 5$ nm reaches about two orders of magnitude better dipole sensitivity due to the proximity enhancement scaling as $1/r^3$. Our calculations suggest that such an NV center 5 nm below the surface has the sensitivity to detect a single electron spin within ~15 nm of the surface with only 1 s of averaging. By averaging for 1 minute, it should be possible to detect a single proton spin on the surface. The sensitivity could be further improved by increasing the photon collection efficiency[7] and by extending the spin coherence time via dynamical decoupling techniques.[43-45]

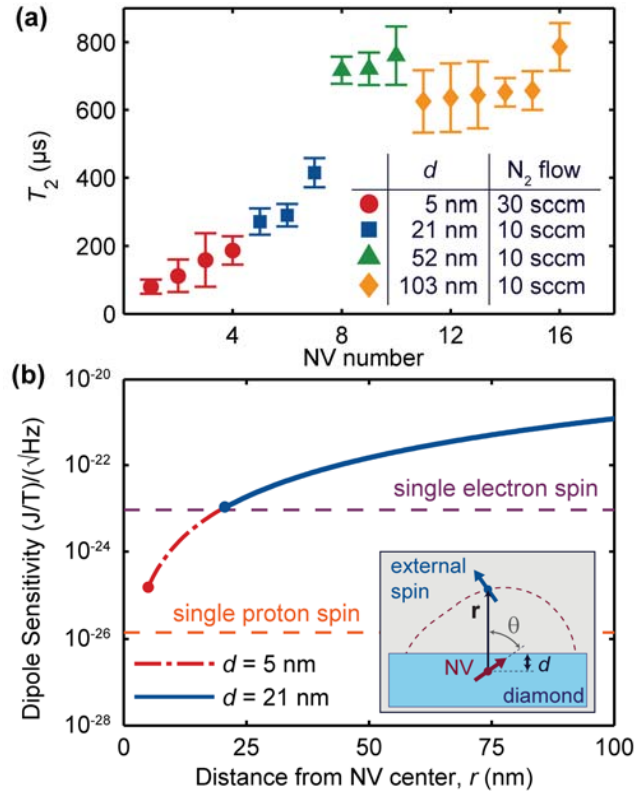

**Figure 4:** (Color) (a) Summary of $T_2$ measurements for doped NV centers. (b) Predicted magnetic dipole sensitivity as a function of distance from the doped NV center, $r$, for the highest-coherence NV center in the $d = 5$ (red) and 21 nm (blue) samples. We assume **r** ∥ [100], such that $d$ is the minimum separation. The required sensitivities for detecting single electron and proton spins with 1s averaging are shown as dashed lines. Inset: schematic of a doped NV center probe (symmetry axis [111]) in the diamond sample and an external spin at displacement **r**. The dashed curve is an iso-sensitivity contour,[29] assuming the external spin is oriented to optimize the sensitivity for each displacement angle, $\theta$.

In summary, we have demonstrated a growth technique of nanometer-scale, depth-controlled nitrogen delta-doping for applications requiring shallow NV centers. The artificially created NV centers were verified to be confined within ± 4 nm and showed consistently long $T_2 > 100$ (600) μs at 5 (52) nm below the surface. This technique provides a powerful approach to materials engineering for atomic-scale sensing and hybrid quantum information technologies by using NV centers to interface with degrees of freedom external to the diamond crystal.

This work was supported by DARPA and AFOSR. We are grateful to D. Rugar and J. Mamin for insightful discussion, and thank B. B. Buckley, C. G. Yale, D. J. Christle and S. J. Brown for experimental assistance and helpful discussions.

# Supplemental Material for
# "Engineering shallow spins in diamond with nitrogen delta-doping"

Kenichi Ohno, F. Joseph Heremans, Lee C. Bassett, Bryan A. Myers, David M. Toyli,

Ania C. Bleszynski Jayich, Christopher J. Palmstrøm, and David D. Awschalom

*Center for Spintronics and Quantum Computation,*

*University of California, Santa Barbara, California 93106, USA*

1. **Nitrogen residual gas analysis**

All the samples in this study were grown by a Seki Technotron AX6300 1.5 kW microwave plasma-enhanced chemical vapor deposition (PE-CVD) system. In order to estimate the doping gas residual time, a limiting factor in the sharpness of our nitrogen doped layer interface, we performed residual gas analysis (RGA) of nitrogen in the growth chamber. The RGA probe was directly inserted into the CVD growth chamber via a port. We introduced $N_2$ gas at fixed flow rates of 10, 50 and 100 sccm while keeping all other gas conditions similar to the actual growth parameters ($H_2$ flow of 400 sccm and the chamber pressure of 25 Torr) and measured the nitrogen partial pressure decay with the RGA after the nitrogen flow was cut off, shown in Fig. S1. These data were fit to a single exponential decay resulting in a gas decay time constant of $19.8 \pm 1.7$ sec, which is consistent over the three different nitrogen flows. Given a growth rate of $8.2 \pm 3.2$ nm / hour, this decay time corresponds to a thickness of less than 1 Å during this transition period in which the nitrogen flow was cut off.

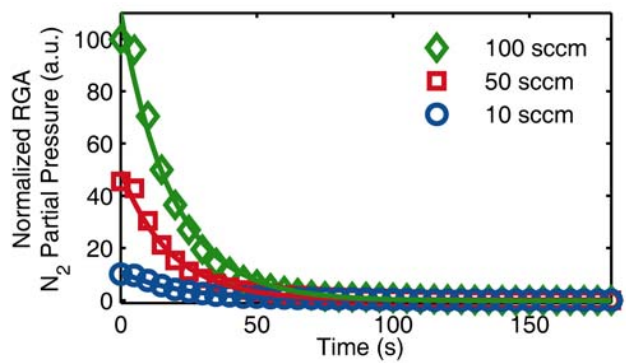

**Figure S1:** Normalized partial pressure of nitrogen gas measured with RGA as a function of time for 100, 50, and 10 sccm $N_2$ flow.

## 2. Nitrogen doping calibration

Figure S2 shows a $^{14}$N secondary ion mass spectrometry (SIMS) measurement of a grown sample that contains four nitrogen doped layers. We grew a CVD diamond film with our standard growth conditions on top of a type Ib high-pressure high-temperature (100) diamond substrate. We used $CH_4$ and $N_2$ gases of natural isotopic abundance for this sample. For each nitrogen flow rate (50, 20, 10, and 5 sccm from deepest to shallowest), we introduce the $N_2$ gas for 1 hour and separate each doped layer with 2 hours of undoped diamond growth. By fitting the SIMS data, we identify three peaks with nitrogen concentrations of $(3.1 \pm 0.3) \times 10^{16}$ cm$^{-3}$, $(1.3 \pm 0.3) \times 10^{16}$ cm$^{-3}$, and $(0.8 \pm 0.6) \times 10^{16}$ cm$^{-3}$. From the relative intensity and the doping order, we attribute these peaks to the doped nitrogen concentration resulting from the 50, 20, 10 sccm $N_2$ flow rates, suggesting that the 5 sccm $N_2$ flow rate results in a nitrogen doping concentration that is below the noise floor of $0.6 \times 10^{16}$ cm$^{-3}$.

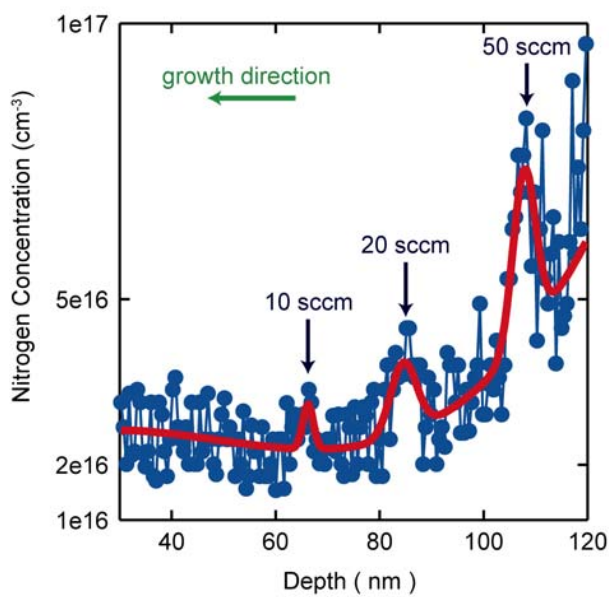

**Figure S2:** $^{14}$N SIMS scan of a CVD-grown diamond film with multiple nitrogen doped layers with varying $N_2$ flow. The red line is a three Gaussian fit on top of a cubic polynomial to account for the fluctuations of the background.

### 3. Hahn echo analysis

In order to probe the spin environment and to extract spin coherence times ($T_2$) of doped NV centers, we used a Hahn echo technique consisting of a $\pi/2 - \tau - \pi - \tau - \pi/2$ sequence for the pulsed microwave field, illustrated schematically in Figure S3(a). The sequence begins by initializing the NV center spin into $m_s$ = 0 state with a green (532 nm) laser, after which a first $\pi/2$ pulse rotates the NV center spin to the superposition of $m_s$ = 0 and -1 states. The center $\pi$ pulse mitigates the effects from field fluctuations slower than $\tau$ through cancellation of phase accumulated before and after the pulse, thus restoring the spin coherence. The final $\pi/2$ pulse projects the NV center spin back into the basis where spin-dependent photoluminescence readout can be performed; thus coherence is mapped onto a population difference between the $m_s$ = 0 and - 1 states.

The Hahn echo signal, measuring the probability $p(\tau)$ that the electron spin is in the $m_s = 0$ spin sublevel at the end of the sequence with total delay $2\tau$, can be written in the form

$$p(\tau) = \frac{1+S(\tau)}{2}, \qquad (S3.1)$$

where $S(\tau) \in [0,1]$ measures the coherence of the electron spin. In the case of $^{15}$NV centers [S1] coupled only to their host $^{15}$N nuclear spin,

$$S(\tau) = S_{^{15}N}(\tau) D(\tau) \qquad (S3.2)$$

is the product of a term describing the coherent interaction with the $^{15}$N spin, $S_{^{15}N}(\tau)$, and the dephasing term,

$$D(\tau) = \exp\left[-\left(\frac{2\tau}{T_2}\right)^n\right], \qquad (S3.3)$$

Where $T_2$ is the spin coherence time and the $n$ is the power index which is dependent on the decoherence mechanisms and left as a fitting parameter. The electron spin echo envelope modulation (ESEEM) function $S_{^{15}N}(\tau)$ can be written as

$$S_{^{15}N}(\tau) = 1 - 2C \sin^2(\pi f_0^{^{15}N} \tau) \sin^2(\pi f_{-1}^{^{15}N} \tau) \qquad (S3.4)$$

where the contrast $C$ depends on the external magnetic field vector and $f_{m_s}^{^{15}N}$ is the Larmor precession frequency of the $^{15}$N nuclear spin depending on the NV center spin state $m_s$. At the relatively low magnetic fields used in our experiments, the faster frequency component $f_{-1}^{^{15}N} \approx 3$MHz is dominated by the $^{15}$N hyperfine frequency, while the slower frequency component $f_0^{^{15}N}$ is proportional to the magnetic field component perpendicular to the NV center axis. According to equation (S3.4), $S_{^{15}N}(\tau) = 1$ when $\tau$ is a multiple of $1/f_0^{^{15}N}$. Therefore, when we measure the Hahn echo signal in discrete steps of $1/f_0^{^{15}N}$ (hereafter referred to as 'nodes'), the Hahn echo signal gives us a direct measurement of the $T_2$ coherence decay $D(\tau)$.

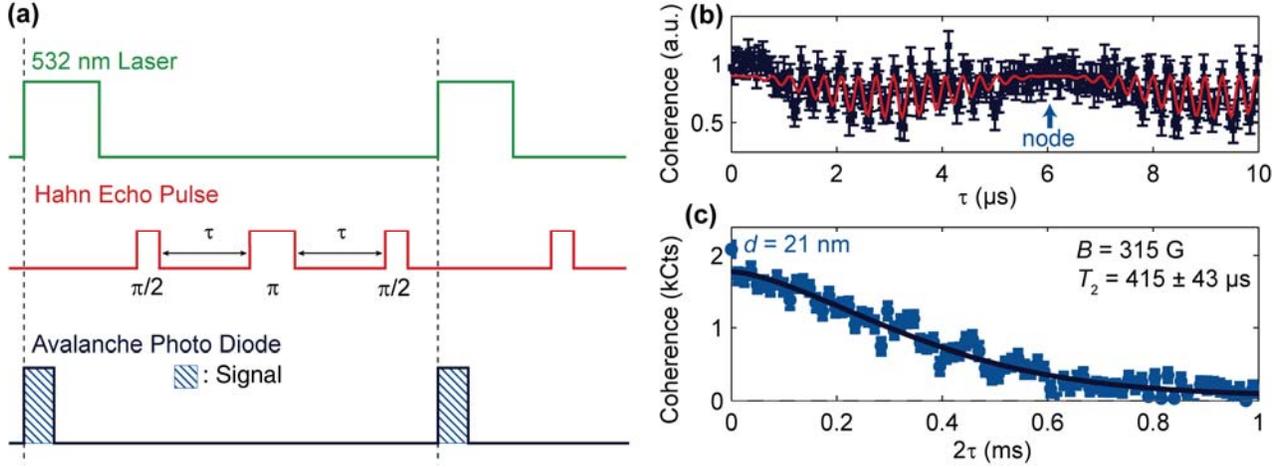

**Figure S3:** (a) Hahn echo pulse sequence, (b) Spin coherence measured in $d$ = 21 nm by Hahn echo sequence for $\tau$ = 0 - 10 μs in 25 ns steps. The red curve is the fit to equation (S3.4). (c) Hahn echo decay curve of the same NV center measured in $2\tau$ = 12.34 μs data steps. The black curve is the fit to equation (S3.3).

Figure S3(b) shows the Hahn echo signal at short times for a doped NV center in the $d$ = 21 nm sample. The red curve shows the fit according to equation (S3.4), resulting in two frequency components $f_{-1}^{^{15}N}$ = 3.06 ± 0.01 MHz and $f_0^{^{15}N}$ = 162 ± 8 kHz. We also performed ESEEM measurements from $\tau$ = 50 μs to 60 μs to accurately determine the node spacing for the $T_2$ decay. To extract $T_2$, the Hahn echo measurement was extended up to $2\tau$ > 1 ms in discrete steps of $\tau = 1/f_0^{^{15}N}$, shown in Fig. S3(c) for the $d$ = 21 nm sample. The black curve is the fit to equation (S3.3), revealing $T_2$ = 415 ± 43 μs of this NV center. All measurements of $T_2$ reported in the main text were taken in this same fashion.

## 4. Coupling to $^{13}$C-enriched layers

In the main text we present the results of ESEEM measurements on doped NV centers in a 12C layer of a $^{13}$C/$^{12}$C/$^{13}$C diamond isotopically layered structure (see Fig. 2 in the main text). Of 29 NV centers studied in this structure, only 6 show ESEEM coherence modulations due to coupling to $^{13}$C while the remaining ~ 80 % of NV centers show no $^{13}$C coupling. In this section, we construct a model to estimate how far these NV centers must be from the interface with the $^{13}$C layer such that no $^{13}$C modulation is observed. Given the known thickness of the $^{12}$C layer (15 nm), this minimum separation constrains the depth dispersion of our doped NV centers. At the magnetic field used in our experiments, our calculations predict that any NV centers within ~ 4 nm of either $^{13}$C layer would show measurable coherence modulation in Hahn echo measurements, implying a doped NV dispersion less than ± 3 nm.

At nanometer-scale separations, the interaction between an NV center and individual $^{13}$C nuclear spins is very weak, and is well described by the Hamiltonian ($\hbar = 1$),

$$H = \Delta S_z^2 - \mu_e B_z S_z + (\mu_n \mathbf{B} + S_z \mathbf{A}) \cdot \mathbf{I} \qquad (S4.1)$$

where $\Delta/2\pi$ = 2.87 GHz is the zero-field splitting, $\mathbf{S}(\mathbf{I})$ and $\mu_e$ ($\mu_n$) are the electron (nuclear) spin and magnetic moment respectively, $\mathbf{B}$ is the magnetic field vector, and $\mathbf{A}$ is the hyperfine interaction vector. Each $^{13}$C nucleus produces a modulation in the coherence envelope of a Hahn echo measurement analogous to the $^{15}$N interaction described by equation (S3.4). For the $j^{th}$ nucleus,

$$S_j = 1 - 2C_j \sin^2(\pi f_0^{(j)} \tau) \sin^2(\pi f_{-1}^{(j)} \tau), \qquad (S4.2)$$

where $f_{m_s}$ is the $m_s$-dependent $^{13}$C Larmor precession frequency due to the effective magnetic field seen by the nucleus, $\mathbf{B}_{m_s} = \mathbf{B} + m_s \mathbf{A}/\mu_n$, and

$$C_j = \frac{|\mathbf{B}_0 \times \mathbf{B}_1|^2}{|\mathbf{B}_0|^2 |\mathbf{B}_1|^2}. \qquad (S4.3)$$

Assuming a point-dipole hyperfine interaction, $\mathbf{A}$ is given by

$$\mathbf{A}_{dip}(\mathbf{r}_j) = \frac{\mu_0 \mu_n \mu_e}{4\pi r_j^3} (3\hat{\mathbf{r}} \cos\theta_j - \hat{\mathbf{z}}), \qquad (S4.4)$$

where $\theta_j$ is the angle between the NV center symmetry axis $\hat{\mathbf{z}}$ and the displacement vector $\mathbf{r}_j$ (see Fig. S4). The dipole approximation will be ultimately justified by our conclusion that $r \approx$ 4 nm. Furthermore, at our applied field, $B$ = 136 G, $\mu_n B$ = 145 kHz >> |$\mathbf{A}$| (|$\mathbf{A}$| ≈ 600 Hz at $r \approx$ 4 nm) and can therefore approximate $|\mathbf{B}_0| \approx |\mathbf{B}_1| \approx B$ such that

$$S_j(\tau) \cong 1 - 2C_j \sin^4(\pi f_L \tau), \tag{S4.5}$$

where $f_L = 1.071$ kHz/G is the bare $^{13}$C Larmor precession frequency and

$$C_j \cong \frac{|\mathbf{A}(\mathbf{r}_j)|^2 \sin^2(\alpha_j)}{\mu_n^2 B^2}$$

$$= \left(\frac{\mu_0 \mu_e}{4\pi r_j^3 B}\right)^2 \{3\cos^2\theta_j + 1 - (3\cos\theta_j \hat{\mathbf{r}}_j \cdot \hat{\mathbf{e}} - \hat{\mathbf{z}} \cdot \hat{\mathbf{e}})^2\}, \tag{S4.6}$$

where $\alpha_j$ is the angle between $\mathbf{A}(\mathbf{r}_j)$ and $\mathbf{B} = B\hat{\mathbf{e}}$. If the magnetic field is applied along the NV symmetry axis, $\hat{\mathbf{e}} \parallel \hat{\mathbf{z}}$, then equation (S4.6) simplifies further to

$$C_j = \left(\frac{3\mu_0 \mu_e}{8\pi r_j^3 B}\right)^2 \sin^2(2\theta_j), \tag{S4.7}$$

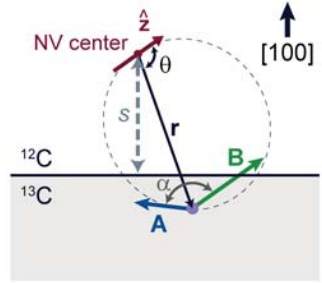

**Figure S4**: Schematic geometry of NV center and $^{13}$C nuclear spin considered in our model. The NV center located in the $^{12}$C layer is set at the origin with its spin parallel to $\hat{\mathbf{z}} \parallel \langle 111 \rangle$. The $^{13}$C spin is located at $\mathbf{r}$, with an angle $\theta$ relative to $\hat{\mathbf{z}}$. The hyperfine interaction induced at the $^{13}$C, $\mathbf{A}$, makes an angle $\alpha$ with the external magnetic field $\mathbf{B}$, and $s$ is the distance from the NV center to the $^{12}$C/$^{13}$C layer interface.

The total modulation response from an ensemble of spins (neglecting the decoherence decay) is given by $S_{bath}(\tau) = \prod_j (1 - 2C_j) \cong 1 - 2\sum_j C_j$ to lowest order in the $C_j \ll 1$. Hence we can approximate $S_{bath}(\tau)$ by the integral

$$S_{bath}(\tau) \cong 1 - 2\int dr^3 n(\mathbf{r}) C(\mathbf{r}), \tag{S4.8}$$

where $n(\mathbf{r})$ is the $^{13}$C density distribution.

For an isotropic distribution of $^{13}$C spins with concentration $x$, i.e., $n(\mathbf{r}) = xn_0$, where $n_0 = 176$ nm$^{-3}$ is the diamond atomic density, we can use equation (S4.8) to calculate the critical coupling radius, $r_c$, beyond which the integrated effect of $^{13}$C spins is negligible. With $C(r,\theta) = c_0 \sin^2(2\theta_j)/r^6$, the total contrast from such a shell of bath spins with $r > r_0$ is

$$C_{shell}(r_0) = 2\pi c_0 xn_0 \int_0^\pi d\theta \sin\theta \sin^2(2\theta) \int_{r_0}^\infty \frac{dr}{r^4}$$

$$= \frac{32\pi c_0 xn_0}{45 r_0^3}, \qquad (S4.9)$$

from which we define

$$r_c = \left( \frac{32\pi c_0 xn_0}{45 C_{min}} \right)^{1/3}, \qquad (S4.10)$$

where $C_{min}$ is the minimum observable contrast. At the magnetic field $B = 136$ G applied in our experiments, $c_0 = \left( \frac{3\mu_0 \mu_e}{8\pi B} \right)^2 = 0.042$ nm$^6$. In a sample with a natural abundance of $^{13}$C ($x_{nat} = 0.011$), and assuming $C_{min} \approx 10\%$, equation (S4.10) implies $r_c = 1.2$ nm, which is consistent with previous studies.[S2] In $^{13}$C-enriched diamond, however, $r_c$ is increased by the factor $(x/x_{nat})^{1/3}$, up to 5.5 nm for $x = 1$.

In a similar fashion, we can estimate the minimum distance $s$ that an NV center embedded in the $^{12}$C layer at the center of our `isotopic sandwich' structure can be separated from one of the $^{13}$C-layer interfaces and still show no $^{13}$C modulation. For simplicity, and since the coupling in the limiting case we seek will be dominated by the closest $^{13}$C spins to the NV center, we approximate $C(r,\theta) \cong C(r,\theta_0)$, where $\theta_0 = 55°$ is the angle between the NV center symmetry axis (<111> family) and the [100] vector normal to the layers. Then, choosing a spherical coordinate system $(r,\theta',\phi')$ where $\theta'$ is relative to the [100] axis, the integral of equation (S4.8) becomes

$$C_{layer}(s) \cong c_0 xn_0 \sin^2(2\theta_0) \int d\phi' \int_s^\infty \frac{dr}{r^4} \int_0^{\cos^{-1}(\frac{s}{r})} d\theta' \sin\theta'$$

$$= \frac{\pi c_0 xn_0}{6 s^3} \sin^2(2\theta_0), \qquad (S4.11)$$

Again, assuming a minimum observable contrast of 10 % and using our SIMS-measured $^{13}$C concentration, $x = 0.8$, we obtain a maximum separation $s_{max} = 3.0$ nm. This implies that the ~ 80% of NV centers that show no $^{13}$C coupling in our measurements must be more than ~ 3 nm from the interface with either $^{13}$C layer. Assuming a Gaussian distribution for doped NV centers centered in the 15-nm-thick $^{12}$C layer, this implies a depth dispersion

$$\sigma = \frac{(7.5-3.0)\,\text{nm}}{\sqrt{2}\,\text{erf}^{-1}(23/29)} = 3.6\,\text{nm}. \tag{S4.12}$$

As discussed in the main text, it is possible that the doped NV centers that do show $^{13}$C modulation may be coupled to residual $^{13}$C impurities in the $^{12}$C layer rather than to the interfaces, suggesting the dispersion is consistent with the 2-nm layer thickness inferred from the growth rate.

### 5. AC sensitivity analysis

Here, we explain how we calculate the ac dipole sensitivity shown in Fig. 4(b). We assume that the target dipole can be initially polarized and driven resonantly such that ac sensing techniques may be used. We followed the method for ac magnetic field sensitivity outlined in Refs. S3 and S4, adapted accordingly to our experimental parameters. The ac magnetic field sensitivity is given by the following equation,

$$\eta_B = \frac{\pi\hbar}{g\mu_B C\sqrt{2\tau}F(\tau)}, \tag{S5.1}$$

where $\hbar$ is the reduced Planck's constant, $g \sim 2$ is the electron $g$-factor, $\mu_B$ is the Bohr magnetron, $C$ is a factor that estimates the photon shot noise, $2\tau$ is the spin-echo interval, and $F(\tau)$ is the amplitude of the envelope function of spin echo signal. We calculate parameters $2\tau$, $F(\tau)$, and $C$ from our experiment.

In order to detect the external ac magnetic field, its period $1/\nu$ should match the spin echo interval, $2\tau$, such that the probe NV center spin accumulates a net phase. We neglect the spin polarization and read out time $\tau_m = 2$ μs in the echo sequence as they are much smaller than the $2\tau$ timescale (usually longer than 100 μs), giving the estimate

$$\tau_m + 2\tau \approx 2\tau = \frac{1}{\nu}. \tag{S5.2}$$

In isotopically pure $^{12}$C diamond, with natural $^{14}$N nuclear spins there are no ESEEM frequency components, in which case the best sensitivity is when $\sqrt{2\tau}F(\tau)$ is maximized. However, for our doped NV centers, the $^{15}$N nuclear spins contribute to the ESEEM signal so we must eliminate this contribution when calculating the sensitivity of detecting external fields. A simple way of getting around this is to choose the external ac frequency, $v$, such that

$$2mv = f_0^{15N}, \qquad (S5.3)$$

where $m$ is an integer and $f_0^{15N}$ is the node frequency described in the section above. In this way, the external field causes constructive phase accumulation while $^{15}N$-induced ESEEM components are cancelled out. At this frequency, $F(\tau)$ contains only a dephasing term as,

$$F(\tau) = \exp\left\{-\left(\frac{2\tau}{T_2}\right)^n\right\}. \qquad (S5.4)$$

The best sensitivity is acquired when $\sqrt{2\tau}F(\tau)$ is maximized. For the doped NV center in sample $d = 5$ nm (shown in Fig. 3(b)), we extract $T_2 = 187$ μs, $n = 2.442$, $f_0^{15N} = 95.4$ kHz and calculate an ac field frequency of $v = 9.54$ kHz.

The $C$ parameter in equation (S5.1) estimates the photon noise of the experiment, and given that our signal read-out window is ~ 400 ns and the average detected photon count is much less than 1, our read-out sensitivity is shot-noise limited. $C$ can be written as

$$C = \left(\sqrt{1 + 2\frac{(a_0 + a_1 + a_0 a_1)}{(a_0 - a_1)^2}}\right)^{-1}, \qquad (S5.5)$$

where $a_0$ and $a_1$ are the photon counts for the $m_s = 0$ and $\pm 1$ states in a single measurement duty cycle.[S3] For the doped NV center discussed above, $a_0 = 0.0056$, $a_1 = 0.0038$ and $C = 0.013$. With all these experimentally determined parameters entered into the equation (S5.1), we get a field sensitivity $\eta_B = 172$ nT/√Hz.

In order to translate the magnetic field sensitivity to a spin sensing application, we estimate the optimum sensitivity to a dipole at displacement $\mathbf{r}$ from a doped NV center. Defining the NV symmetry axis ([111] direction) as $\hat{\mathbf{z}}$ and assuming $\mathbf{r}$ lies in the $x$-$z$ plane making an angle $\theta$ to $\hat{\mathbf{z}}$ (see inset to Fig. 4(b) in the main text), the coordinate of this external spin can be written as $\mathbf{r} = -r\begin{pmatrix}\sin\theta \\ \cos\theta\end{pmatrix}$. The magnetic dipole of this external spin, $\mathbf{m}$, which we assume parallel to the external magnetic field, $\mathbf{B}_{\text{ext}}$, can be written as $\mathbf{m} = m\begin{pmatrix}\sin\alpha \\ \cos\alpha\end{pmatrix}$, where $\alpha$ is the angle between $\mathbf{B}_{\text{ext}}$ and $\hat{\mathbf{z}}$. The dipole field at the NV center induced by this external spin is given by

$$\mathbf{B(r)} = \frac{\mu_o m}{4\pi r^5}\{3(\mathbf{m}\cdot\mathbf{r})\mathbf{r} - \mathbf{m}r^2\}$$

$$= \frac{\mu_o m}{4\pi r^3}\begin{bmatrix}(3\cos\theta\sin\theta)\cos\alpha + (3\sin^2\theta-1)\sin\alpha \\ (3\cos^2\theta-1)\cos\alpha + (3\cos\theta\sin\theta)\sin\alpha\end{bmatrix}. \quad (S5.6)$$

For small values of $B_{ext}$ (relative to the NV center zero field splitting $\Delta/2\pi = 2.87$ GHz), the NV center is sensitive to the projection of the dipole field component along its symmetry axis, given by

$$B_z = \frac{\mu_o m}{4\pi r^3}\{(3\cos^2\theta-1)\cos\alpha + (3\cos\theta\sin\theta)\sin\alpha\}. \quad (S5.7)$$

For fixed $\theta$, $B_z$ takes the maximum value of

$$B_z^{max}(\theta) = \frac{\mu_o m}{4\pi r^3}\sqrt{3\cos^2\theta+1} \quad (S5.8)$$

with $\mathbf{B}_{ext}$ applied at the angle $\alpha_o$ such that $\tan\alpha_o = (3\cos\theta\sin\theta)/(3\cos^2\theta-1)$. Combining (S5.1) and (S5.8) gives the optimum dipole sensitivity,

$$\eta_m(r,\theta) = \frac{\pi\hbar}{g\mu_B C\sqrt{2\tau F(\tau)}}\cdot\frac{4\pi r^3}{\mu_0\sqrt{3\cos^2\theta+1}} \quad (S5.9)$$

In Fig. 4(b), we plot equation (S5.9) as a function of distance from the NV for the doped NV centers in the $d = 5$ and 21 nm samples with the longest spin coherence times, assuming $\mathbf{r}$ perpendicular to the [100] surface such that $\theta = 54.7°$. The inset of Fig. 4(b) also shows the iso-sensitivity contour (dotted line), which is given by

$$r_{iso}(\theta) = \left[\frac{\mu_o m}{4\pi B_o}\sqrt{3\cos^2\theta+1}\right]^{1/3}. \quad (S5.10)$$

We could further improve the dipole sensitivity by enhancing a number of parameters in equation (S5.9). We could increase the shot noise parameter, $C$, up to $\approx 0.05$ by improving the photon collection efficiency via an oil immersion lens and high NA objective.[S3] The value of $\sqrt{2\tau}F(\tau)$ could be increased by extending $T_2$ via dynamical decoupling schemes.[S5-S7]